\documentclass[nanomaterials,article,accept,moreauthors,pdftex]{Definitions/mdpi}

\usepackage{upgreek}
\firstpage{1}
\pubvolume{11}
\issuenum{3}
\articlenumber{675}
\pubyear{2021}
\copyrightyear{2021}
\externaleditor{Academic Editor: Andrey B. Evlyukhin} % For journal Automation, please change Academic Editor to "Communicated by" %MDPI: Please check whether there nned to add academic editor
\datereceived{3 February 2021}
\dateaccepted{5 March 2021}
\datepublished{9 March 2021}
\hreflink{https://doi.org/10.3390/nano\linebreak 11030675} % If needed use \linebreak
\pdfoutput=1

\Title{Anisotropy of the In-Plane and Out-of-Plane Resistivity and the Hall Effect in the Normal State of Vicinal-Grown YBa$_{2}$Cu$_{3}$O$_{7-\delta}$ Thin Films}

\TitleCitation{Anisotropy of the In-Plane and Out-of-Plane Resistivity and the Hall Effect in the Normal State of Vicinal-Grown YBa$_{2}$Cu$_{3}$O$_{7-\delta}$ Thin Films}

\Author{Gernot Heine $^{1}$, Wolfgang Lang $^{1,}$*\href{https://orcid.org/0000-0001-8722-2674}{\orcidicon}, Roman R\"ossler $^{2}$ and Johannes D. Pedarnig $^{2}$\href{https://orcid.org/0000-0002-7842-3922}{\orcidicon}} %MDPI: Please carefully check the accuracy of names and affiliations.

\AuthorNames{Gernot Heine, Wolfgang Lang, Roman R\"ossler and Johannes D. Pedarnig}

\AuthorCitation{Heine, G.; Lang, W.; R\"ossler, R.; Pedarnig, J.D.}

\address{%
$^{1}$ \quad Faculty of Physics, University of Vienna, Boltzmanngasse 5, \mbox{A-1090 Wien, Austria}; Gernot.Heine@bev.gv.at\\
$^{2}$ \quad Institute of Applied Physics, Johannes-Kepler-University Linz, Altenbergerstrasse 69,  A-4040 Linz, Austria; Roman.Roessler@voestalpine.com (R.R.); Johannes.Pedarnig@jku.at (J.D.P.)}

\corres{Correspondence: wolfgang.lang@univie.ac.at}
\abstract{The resistivity and the Hall effect in the copper-oxide high-temperature superconductor YBa$_{2}$Cu$_{3}$O$_{7-\delta}$ (YBCO) are remarkably anisotropic. Using a thin film of YBCO grown on an off-axis cut SrTiO$_3$ substrate allows one to investigate these anisotropic transport properties in a planar and well-defined sample geometry employing a homogeneous current density. In the normal state, the Hall voltage probed parallel to the copper-oxide layers is positive and strongly temperature dependent, whereas the out-of-plane Hall voltage is negative and almost temperature independent. The results confirm previous measurements on single crystals by an entirely different measurement method and demonstrate that vicinal thin films might be also useful for investigations of other layered nanomaterials.}


\begin{document}

\section{Introduction}

\textls[-15]{The electrical transport properties of the copper-oxide high-$T_c$ superconductors (HTSCs)} are strongly anisotropic because of their layered crystalline structure. Notably the Hall effect in these materials is one of the still unresolved fundamental puzzles. In the prototypical compound YBa$_{2}$Cu$_{3}$O$_{7-\delta}$ (YBCO) the Hall coefficient in the normal state changes in an unconventional manner by rotating the magnetic field. With the current injected along the CuO$_2$ layers and the magnetic field oriented perpendicular to the layers, a positive (hole-like) and strongly temperature dependent Hall effect is observed \cite{TOZE87}, but it is negative (electron-like) and temperature-independent over a wide range \cite{HARR92} when the magnetic field is oriented parallel to the layers and the Hall voltage is measured perpendicular to them. In both geometries, the magnetic field is oriented perpendicular to the current to maximize the Lorentz force on the charge carriers and the Hall field is probed orthogonally to both the current and magnetic field.

The Montgomery method \cite{MONT71} is commonly used to determine anisotropic electrical transport properties. However, with tiny single crystals, imperfections of the sample shape and the sizeable area of the contacts, as compared to the distance between them, introduce considerable uncertainty. Also, the density of the injected current varies strongly throughout the sample volume, which might be adverse for some investigations. Therefore, it would be of paramount importance to test and compare the anisotropic transport properties in an entirely different geometry.

The discovery of a huge Seebeck effect in YBa$_{2}$Cu$_{3}$O$_{7-\delta}$ films \cite{LENG92} has led to further investigations of the experimental opportunities of thin films grown on vicinal-cut substrates. The polished surface of a SrTiO$_3$ (STO) substrate, cut at a tilt angle $\alpha$ off the (001) plane, shows a periodic nanoscale step structure. A subsequently deposited YBCO films replicates the surface steps and grows epitaxially in a self-organized roof-tile manner. For tilt angles $\alpha \le 20^\circ$, a regular vicinal growth of the YBCO film and an approximately linear increase of the thermoelectric signal with $\alpha$ has been reported \cite{LENG94}.

The morphology and defect microstructure of such vicinal YBCO films have been investigated in detail and the deviations from the idealized picture discussed \cite{HAAG97,MECH98,MAUR03}. High-resolution electron microscopy has revealed a bending of the YBCO lattice near the STO interface and the relaxation of these defects with increasing distance to the substrate. Correspondingly, the widths of x-ray diffraction rocking curves increase substantially in films with a thickness smaller than 200\,nm \cite{PEDA02}.

Vicinal YBCO films show strong vortex pinning forces and an anisotropic critical current density \cite{JOOS99,EMER04} due to their parallel oriented planar defects. They are also useful for the deconvolution of angle-dependent critical current effects \cite{DURR04}. The reduced symmetry of vicinal films  allows the observation of vortex channeling effects \cite{BERG97} when the magnetic field is parallel to the CuO$_2$ planes, which is not possible in $c$-axis oriented thin films. Another advantage of thin vicinal films is the access to out-of-plane transport properties in situations where the limited penetration depth of optical \cite{MARK97} or light ion irradiation \cite{LANG09} precludes the use of single crystals.

The anisotropic resistivities of several HTSCs thin films, grown on vicinal substrates, have been studied. Among them are optimally doped {\cite{HAAG97}} and oxygen deficient \mbox{YBCO {\cite{HEIN99c}}}, Bi$_2$Sr$_2$CaCu$_2$O$_8$ {\cite{LI96c}}, Hg$_{1-x}$Re$_x$Ba$_2$CaCu$_2$O$_{6+\delta}$ {\cite{YUN00}}, and, recently, the iron-based superconductors Fe$_{1+\delta}$Se$_{0.5}$Te$_{0.5}$ {\cite{BRYJ17}} and NdFeAs(O,F) {\cite{IIDA20}}.

In this paper, we explore the anisotropic transport properties of a vicinal YBCO film and demonstrate that not only anisotropic resistivities but also the in-plane and out-of-plane Hall effect can be measured. Our results are supported by their good accordance with previous findings on single crystals.

%%%%%%%%%%%%%%%%%%%%%%%%%%%%%%%%%%%%%%%%%%
\section{Materials and Methods}

\subsection{Sample Preparation}

Vicinal YBCO films are fabricated by pulsed-laser deposition (PLD) on off-axis cut $\rm{SrTiO}_3$ substrates (TBL Kelpin, Neuhausen, Germany). One side face of the substrate is cut along the [100]  direction but the surface normal is inclined at an angle $\alpha = 10^\circ$ towards the [010] direction. The polished substrates have a rms surface roughness of less than 0.2~nm. A $(200 \pm 20)$\,nm thick YBCO film is grown by PLD with a KrF-excimer laser (LPX 305i, Coherent Inc., Santa Clara, CA, USA, Lambda Physik, Germany, $\lambda = 248\,\rm{nm}$, fluence $3.25\,\rm{J/cm^2}$, pulse duration $20 \,\rm{ns}$, repetition rate $10\,\rm{Hz}$) from a stoichiometric YBCO
target.

\begin{figure}[H]
\includegraphics*[width=0.6\columnwidth]{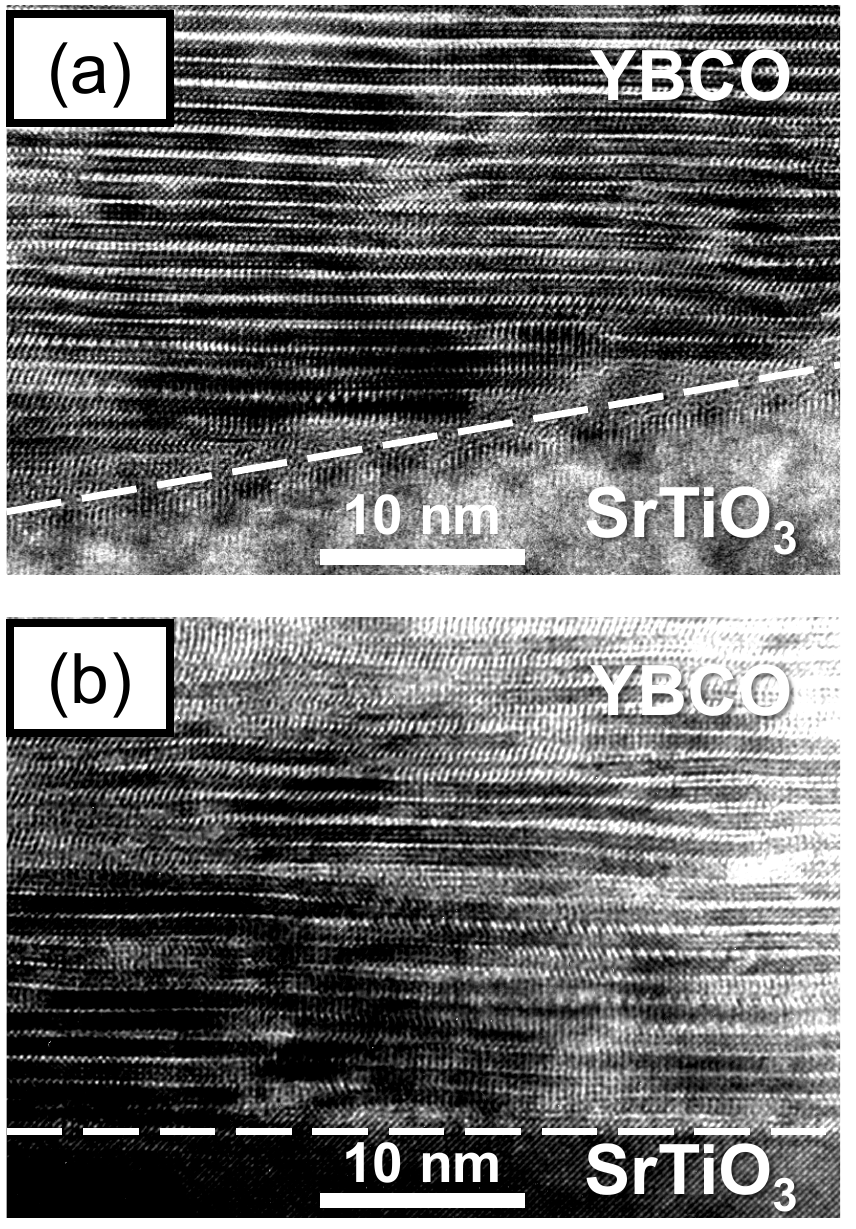}
\caption[]{High-resolution transmission electron microscopy images of vicinal YBCO thin films near the substrate interface. (\textbf{a}) In a cut perpendicular to the vicinal steps, the crystal lattice of YBCO is inclined by $\sim$10$^\circ$ relative to the substrate surface (indicated by a dashed line). (\textbf{b}) In a cut along the vicinal steps the YBCO lattice is oriented parallel to the substrate surface. Adapted from \cite{PEDA02}.}
\label{fig:HRTEM}
\end{figure}

A high-resolution transmission electron microscopy (HRTEM) investigation of the interface microstructure of a similar vicinal YBCO thin film is shown in Figure~\ref{fig:HRTEM} for two different cuts \cite{PEDA02}. The interface between the $\rm{SrTiO}_3$ substrate and the YBCO film is marked by white broken lines. In the cut perpendicular to the vicinal steps, a terraced surface of the substrate is evident in Figure~\ref{fig:HRTEM}a. The YBCO lattice is inclined by $\sim$10$^\circ$ against the interface and exhibits a moderate density of stacking faults that relax with increasing distance to the interface. On the contrary, in a cut along the substrate terraces, the YBCO lattice is oriented parallel to the interface, as can be seen in Figure~\ref{fig:HRTEM}b.

The growth morphology of vicinal YBCO films is confirmed by comparing the X-ray diffraction (XRD) spectra (D8 GADDS Bruker Corp., Billerica, MA, USA) of a reference film on (100) $\rm{SrTiO}_3$ substrate with that of a vicinal-grown YBCO film. The YBCO (00l) diffraction peaks of both samples coincide when the vicinal substrate is tilted by $10^\circ$ in the XRD chamber so that both samples are oriented identically with regard to the CuO$_2$ planes of YBCO \cite{MARK97}. Optical microscopy shows an even surface of our vicinal YBCO film, comparable to the surface structure of $c$-axis oriented films. However, HRTEM investigations of our similarly fabricated vicinal films reveal terrace steps of $18 \pm 6\,$nm height due to the bunching of several unit cells to a larger step {\cite{PEDA02}}.

After deposition, the film is patterned by standard photolithography and wet chemical etching into a cross-shaped geometry (see the inset in Figure~3) with two perpendicular bridges with length $l = 2.3$\,mm and width $w = 0.1$\,mm. Note that one bridge (A---B) is oriented parallel, the other (C---D) perpendicular to the terraces of the vicinal substrate. Finally, eight Au/Pd contact pads are evaporated for the electrical contacts and connected with Au wires. 

\subsection{Transport Measurements}

The electrical measurements are performed in a closed-cycle cryocooler mounted in a continuously rotatable electromagnet supplying a magnetic field of 1~T. A precise adjustment of the applied magnetic field orthogonal and parallel to the CuO$_2$ layers of the vicinal-grown YBCO film is achieved by monitoring the resistance at the low-temperature side of the superconducting transition as a function of the magnet rotation angle. During an extremely slow bidirectional temperature sweep ($\le$ 0.02\,K/min) the data are collected at both polarities of the magnetic field by computer-controlled data acquisition using constant-amplitude ac current and a lock-in amplifier. The temperature was measured by a calibrated platinum resistor in zero magnetic field and controlled by a Cernox resistor with negligible magnetoresistance \cite{HEIN98}.

\subsection{Calculation of the Anisotropic Transport Properties}

\begin{figure}[H]
\includegraphics*[width=0.7\columnwidth]{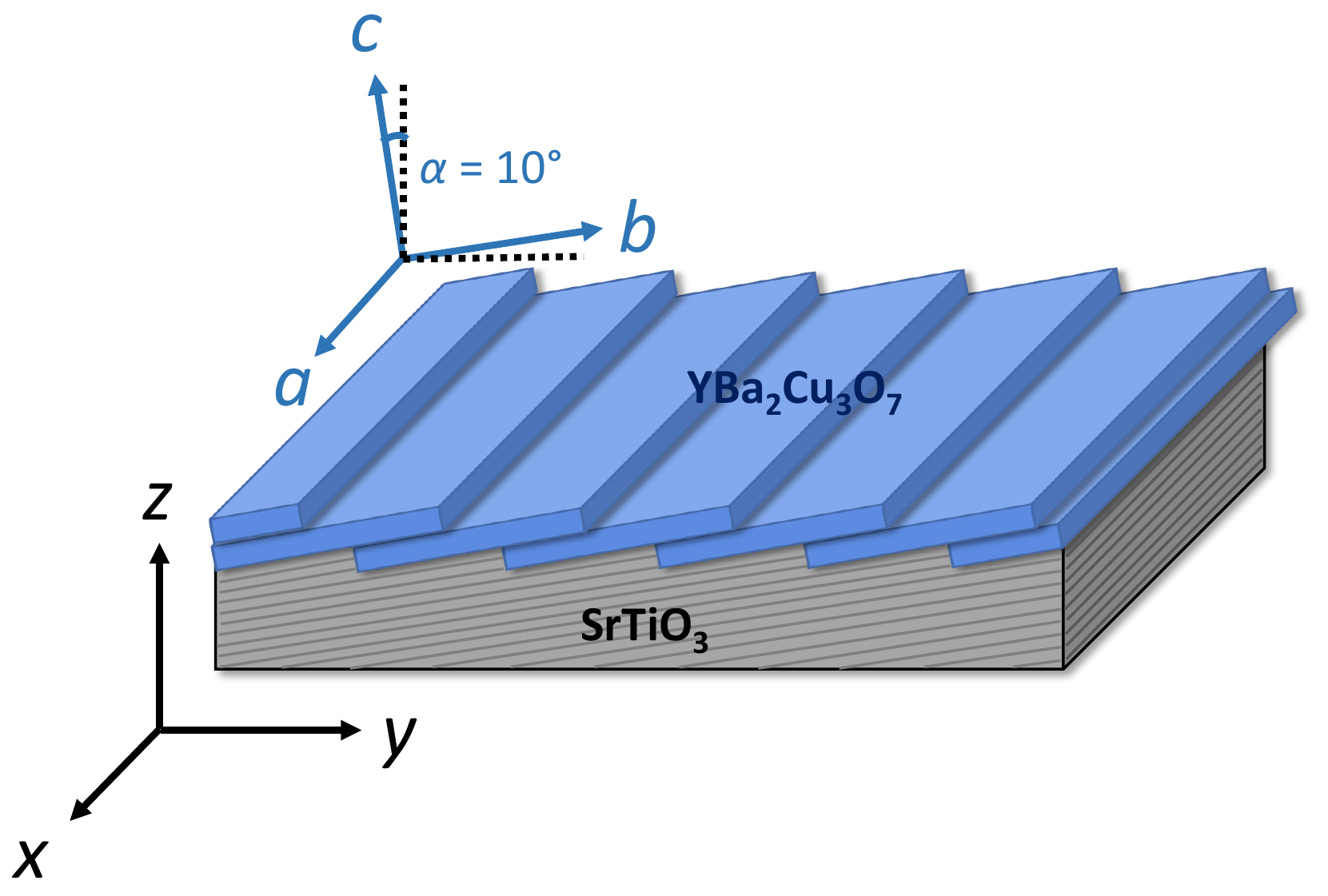}
\caption[]{Substrate and YBCO coordinate systems in thin films grown on vicinal-cut substrates. The two coordinate systems are rotated by an angle $\alpha$ around the $x$ axis.}
\label{fig:coord}
\end{figure}

To evaluate the anisotropic transport parameters in vicinal films, two coordinate systems are introduced, as sketched in Figure~\ref{fig:coord}. The orthogonal axes of the laboratory (substrate) system are labeled $\{x, y, z\}$ and are parallel to the substrate edges, while the crystallographic axes of YBCO are labeled $\{a, b, c\}$. Only the resistivity tensor $\rho_{xyz}$ in the substrate system is experimentally accessible and is connected to the resistivity tensor $\rho_{abc}$ of the YBCO film via the rotational transformation matrix $R$ of the coordinates

\begin{equation} \label{eq:rot_coord}
\rho_{xyz} = R\,\rho _{abc}\, R^{-1},
\end{equation}
where in our vicinal film, $a \parallel x$ and $b \bot c$ are rotated by the vicinal angle $\alpha$ around the \mbox{$x$-axis, hence}

\begin{equation} \label{eq:rot_matrices}
R_x(\alpha)=\begin{pmatrix}
1 & 0 & 0 \\
0 & \cos \alpha & -\sin \alpha \\
0 & \sin \alpha & \cos \alpha \\
\end{pmatrix},
\qquad
R_x^{-1}(\alpha)=\begin{pmatrix}
1 & 0 & 0 \\
0 & \cos \alpha & \sin \alpha \\
0 & -\sin \alpha & \cos \alpha \\
\end{pmatrix}.
\end{equation}

The resistivity tensors in the substrate and the sample systems are, respectively,

\begin{equation} \label{eq:rho_tensors}
\rho _{xyz}=\begin{pmatrix}
\rho_{xx}&\rho_{xy}&\rho_{xz}\\
\rho_{yx}&\rho_{yy}&\rho_{yz}\\
\rho_{zx}&\rho _{zy}&\rho _{zz}\\
\end{pmatrix},
\qquad
\rho _{abc}=\begin{pmatrix}
\rho_{aa}&-\rho_{ba}&-\rho_{ca}\\
\rho_{ba}&\rho_{aa}&-\rho_{cb}\\
\rho_{ca}&\rho _{cb}&\rho _{cc}\\
\end{pmatrix},
\end{equation}
where the latter has been simplified by considering the
tensor symmetry $\rho_{ik} = -\rho_{ki}, i \ne k \in \{a,b,c\}$ according to the Onsager reciprocity relationship \cite{ONSA31}. In materials with tetragonal crystalline symmetry $\rho_{bb} \simeq \rho_{aa}$. Although YBCO has orthogonal symmetry, the twinning in our YBCO films leads to a quasi-tetragonal behavior at the length scales of transport measurements. The relevant components of $\rho_{xyz}$ are then

\begin{align}
\rho_{xx} &= \rho_{aa},  \label{eq:rho_xx} \\
\rho_{yy} &= \rho_{aa} \cos^2 \alpha + \rho_{cc} \sin^2 \alpha, \label{eq:rho_yy} \\
\rho_{yx} = -\rho_{xy} &= \rho_{ba} \cos \alpha - \rho_{ca}\sin \alpha,  \label{eq:rho_yx} \\
\rho_{zx} = -\rho_{xz} &= \rho_{ba} \sin \alpha + \rho_{ca}\cos \alpha, \label{eq:rho_zx} \\
\rho_{yz} = -\rho_{zy}  &= \rho_{aa}\sin \alpha \cos \alpha - \rho_{cc}\sin \alpha \cos \alpha -\rho_{cb}. \label{eq:rho_yz}
\end{align}

Several conditions are essential for reliable measurements of vicinal films. The required high level of crystallinity and epitaxial growth of the films sets an upper limit for the  vicinal angle $\alpha \le 20^\circ$ in YBCO {\cite{LENG94}}. Conversely, $\alpha$ should not be too small to allow for an adequate sensitivity of the measurements. Then, the various elements of the resistivity tensor can be determined by applying a current density $j_k$, in selected directions and measuring the resulting electric field component $E_i$ according to Ohm's law

\begin{equation} \label{eq:ohms_law}
E_i = \rho_{ik} j_k,\qquad i, k \in \{x,y,z\}.
\end{equation}

\subsubsection{Evaluation of Diagonal Resistivity Tensor Elements}

When a current density $j_x$ is injected via the contacts A and B, the resulting electric field probed between contacts A' and B' follows from Equations~(\ref{eq:rho_xx}) and (\ref{eq:ohms_law})

\begin{equation} \label{eq:Ex}
E_x = \rho _{xx}j_x+\underbrace {\rho
_{xy}j_y}_{j_y=0}+\underbrace {\rho _{xz} j_z}_{j_z=0} = \rho_{aa} j_x
\end{equation}
and thus $\rho_{aa} =  \rho _{xx}$.

Alternatively, a current density $j_y$ applied via contacts C and D results in an electric field $E_y$ between contacts C' and D'

\begin{equation} \label{eq:Ey}
E_y = \underbrace {\rho_{yx} j_x}_{j_x=0}+\rho_{yy}j_y+\underbrace {\rho_{yz} j_z}_{j_z=0} = (\rho_{aa}\cos ^2\alpha +\rho _{cc}\sin^2\alpha)j_y
\end{equation}
according to Equations~(\ref{eq:rho_yy}) and (\ref{eq:ohms_law}). By combining Equations~(\ref{eq:Ex}) and (\ref{eq:Ey}) the out-of-plane resistivity of the vicinal film can be calculated from two measurements

\begin{equation} \label{eq:rho_cc}
\rho_{cc} = (\rho_{yy} - \rho_{xx} \cos^2 \alpha)/\sin^2 \alpha.
\end{equation}

\vspace{-6pt}

\subsubsection{Evaluation of off-Diagonal (Hall) Tensor Elements}

Applying a magnetic field $\mathbf{B}$ gives rise to nonzero off-diagonal elements of the resistivity tensors  Equation~(\ref{eq:rho_tensors}) that depend on the magnitude and orientation of $\mathbf{B}$. Here, only the terms to the first order in $\mathbf{B}$ are considered that are evoked by the Lorentz force acting on the charge carriers. For simplicity of the evaluation, Hall effect experiments are commonly designed such that $\mathbf{E} \bot \mathbf{j} \bot \mathbf{B}$. With a substrate surface oriented parallel to the $xy$ plane it would require $\mathbf{B}=(0,0,B_z)$.

However, two arguments suggest aligning $\mathbf{B}$ parallel or perpendicular to the crystallographic $ab$ plane instead. First, the vortex lock-in transition \cite{TACH89,FEIN90} allows one to accurately adjust  $\mathbf{B}$ parallel to the CuO$_2$ planes of YBCO, irrespective of any misorientation of the substrate in the cryostat or uncertainties of the vicinal growth angle $\alpha$. Second, evaluation of the Hall coefficients $R_{ik}^H=\rho_{ik}/B_l, i \neq k \neq l \in \{a,b,c\}$ in the crystallographic system becomes simpler since the tensor components $\rho_{ik}=0$ in a magnetic field $B_l$ such that $i=l$. This reflects the absence of Lorentz force for the current component that is parallel to the magnetic field.

Then, the in-plane Hall coefficient $R_{ba}^H=-R_{ab}^H$ and the out-of-plane Hall coefficient $R_{ca}^H=-R_{ac}^H$ can be determined each with two independent measurements:

\begin{align}
\rho_{yx}=\frac{E_y}{j_x}&=\underbrace{\rho_{ba} \cos \alpha}_{\rho_{ba}=0} - \rho_{ca} \sin \alpha = -R_{ca}^H B_b \sin \alpha, &B_a=B_c=0, \label{eq:RHcaL}\\
\rho_{yx}=\frac{E_y}{j_x}&= \rho_{ba} \cos \alpha - \underbrace{\rho_{ca} \sin \alpha}_{\rho_{ca}=0} = R_{ba}^H B_c \cos \alpha, &B_a=B_b=0, \label{eq:RHbaL}  \\
\rho_{xy}=\frac{E_x}{j_y}&=\rho_{ca} \sin \alpha - \underbrace{\rho_{ba} \cos \alpha}_{\rho_{ba}=0} = R_{ca}^H B_b \sin \alpha, &B_a=B_c=0, \label{eq:RHcaT} \\
\rho_{xy}=\frac{E_x}{j_y}&=\underbrace{\rho_{ca} \sin \alpha}_{\rho_{ca}=0} - \rho_{ba} \cos \alpha = -R_{ba}^H B_c \cos \alpha, &B_a=B_b=0. \label{eq:RHbaT}
\end{align}

%%%%%%%%%%%%%%%%%%%%%%%%%%%%%%%%%%%%%%%%%%
\section{Results and Discussion} \label{sec:results}

\begin{figure}[H]
\includegraphics*[width=0.7\columnwidth]{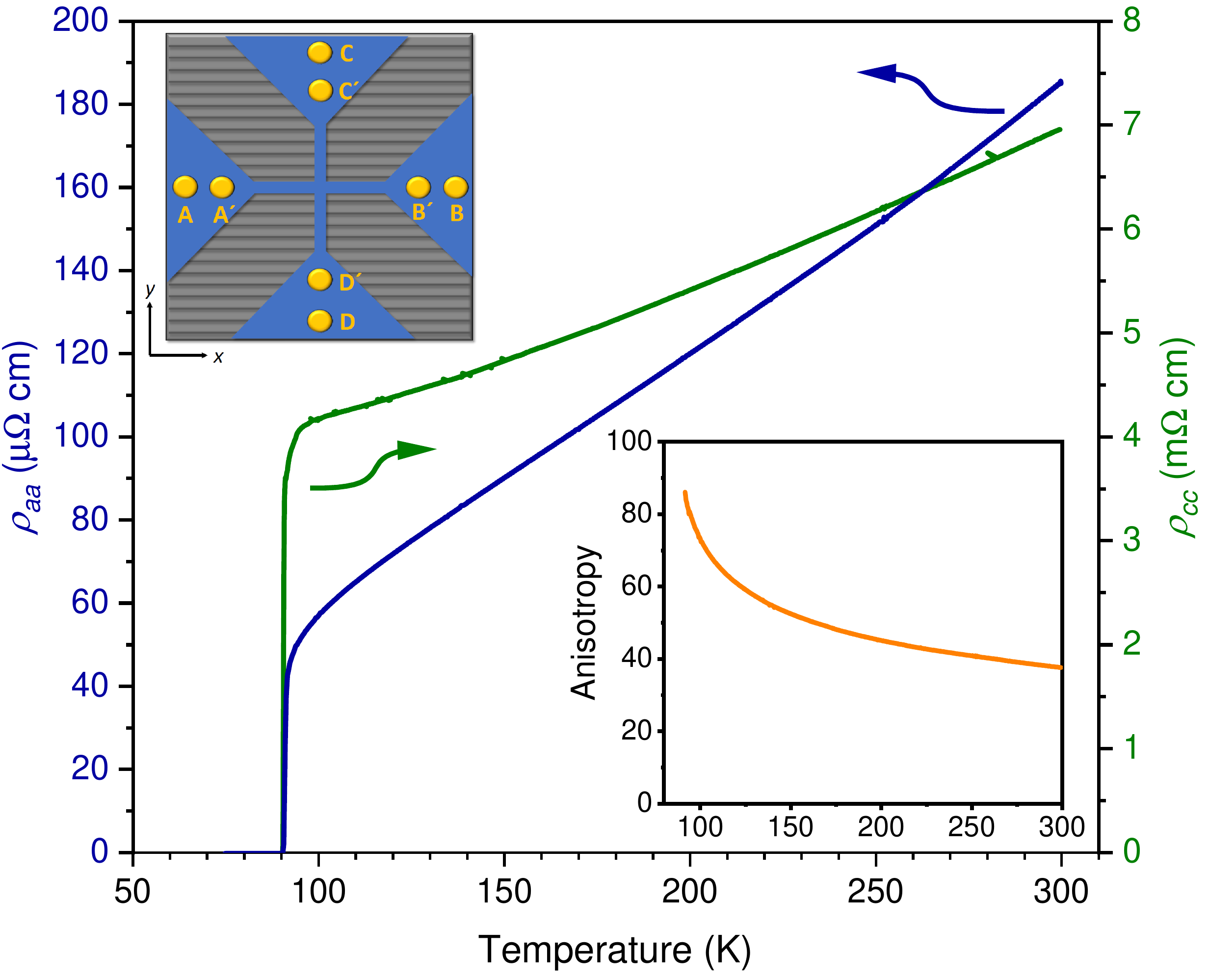}
\caption[]{Temperature dependence of the in-plane resistivity $\rho_{aa}$ (blue line) and out-of-plane resistivity $\rho_{cc}$ (green line) of an optimally doped YBCO thin film grown on a $\alpha = 10^\circ$ vicinal substrate. Bottom right inset: Resistivity anisotropy $\gamma=\rho_{cc}/\rho_{aa}$ as a function of temperature. Top left inset: Schematic view of the cross-shaped patterned YBCO film (blue). Grey lines indicate the terraced structure of the SrTiO$_3$ substrate. Current can be applied in either $x$ direction via contacts (yellow) A---B, or along the $y$ direction via contacts C---D. The neighboring inner contacts (A' and B' or C' and D') are used for resistivity measurements, while for probing the Hall voltage the contacts on the bridge orthogonal to the current flow are used. In parts adapted from {\cite{HEIN99c}}.}
\label{fig:rho}       % Give a unique label
\end{figure}

Figure~\ref{fig:rho} presents the temperature dependencies of the in-plane resistivity $\rho_{aa}$ and the resistivity perpendicular to the CuO$_2$ layers $\rho_{cc}$ of the optimally doped YBCO thin film grown on a $\alpha = 10^\circ$ vicinal SrTiO$_3$ substrate. While $\rho_{aa}=\rho_{xx}$ can be directly determined from one measurement, $\rho_{cc}$ is calculated according to Equation~(\ref{eq:rho_cc}) from two measurements employing current along the $x$ and $y$ axes, respectively. The normal-state resistivity $\rho_{aa}$ has a linear temperature dependence, which extrapolates to an offset $\rho_{aa}(0\,\text{K})\sim -1\, \upmu \Omega$\,cm. At $T \lesssim 110$\,K $\rho_{aa}$ falls below the linear trend because of the paraconductivity stemming from superconducting order-parameter fluctuations \cite{LANG94}.

The out-of-plane normal-state resistivity $\rho_{cc}$ has a metallic temperature dependence, too, in contrast to oxygen-depleted YBCO \cite{ITO91} and HTSCs with stronger anisotropy like Bi$_2$Sr$_2$CaCu$_2$O$_8$ \cite{HEIN99}. The bottom right inset in Figure~\ref{fig:rho} displays the increase of the resistivity anisotropy $\gamma=\rho_{cc}/\rho_{aa}$ from $\gamma(300\,\text{K})=38$ to $\gamma(100\,\text{K})=73$ when reducing the temperature. Both $\rho_{aa}$ and $\rho_{cc}$ vanish at $T_{c0} = 90.3$\,K.

It is instructive to compare the present results obtained in the vicinal film with those in YBCO single crystals and $c$-axis oriented thin films. The in-plane resistivity at room temperature $\rho_{aa}(300\,\text{K})=185\,\upmu \Omega$\,cm is only marginally larger than previously reported for twinned \cite{WINZ91} and untwinned \cite{RICE91} single crystals, if compared to the average value $(\rho_{aa}+\rho_{bb})/2 \simeq 165\,\upmu \Omega$\,cm of the latter. An intercept of the extrapolated normal-state resistivity  $\rho_{aa}(0\,\text{K})$ close to zero or even at slightly negative values of $\rho_{aa}(0\,\text{K})$ is commonly considered a sign of a low density of structural defects \cite{CHIE91a}. Conversely, introducing oxygen disorder enhances the offset, while leaving the slope of $\rho_{aa}(T)$ almost \mbox{unchanged \cite{WANG95b,LANG10R}}.

The out-of-plane resistivity $\rho_{cc}$ of YBCO determined in our vicinal film corresponds intriguingly well with data obtained in twinned single crystals \cite{HOLM93}, both regarding the absolute value and the metallic temperature dependence. It has been proposed that an extended Mott-Ioffe-Regel limit $\rho_{cc}^M \sim 10\, \text{m}\Omega\,$cm bifurcates between HTSCs with metallic behavior when $\rho_{cc}(T) < \rho_{cc}^M$ and semiconducting behavior if $\rho_{cc}(T) > \rho_{cc}^M$ \cite{ITO91}. The latter applies to most of the HTSCs. From Figure~\ref{fig:rho} it is evident that $\rho_{cc}$ falls only marginally below this limit and, indeed, in slightly oxygen-depleted YBCO ($\delta=0.13$) with $\rho_{cc}(300\,\text{K}) \sim \rho_{cc}^M$ a semiconducting behavior is observed. Early experiments in YBCO single crystals also reported somewhat higher $\rho_{cc}(T)$ and a semiconducting behavior below 200\,K \cite{TOZE87,PENN88}. Hence, the results for both out-of-plane and in-plane resistivities in our vicinal films are well comparable to those in optimally doped YBCO single crystals.

\begin{figure}[H]
\includegraphics*[width=0.7\columnwidth]{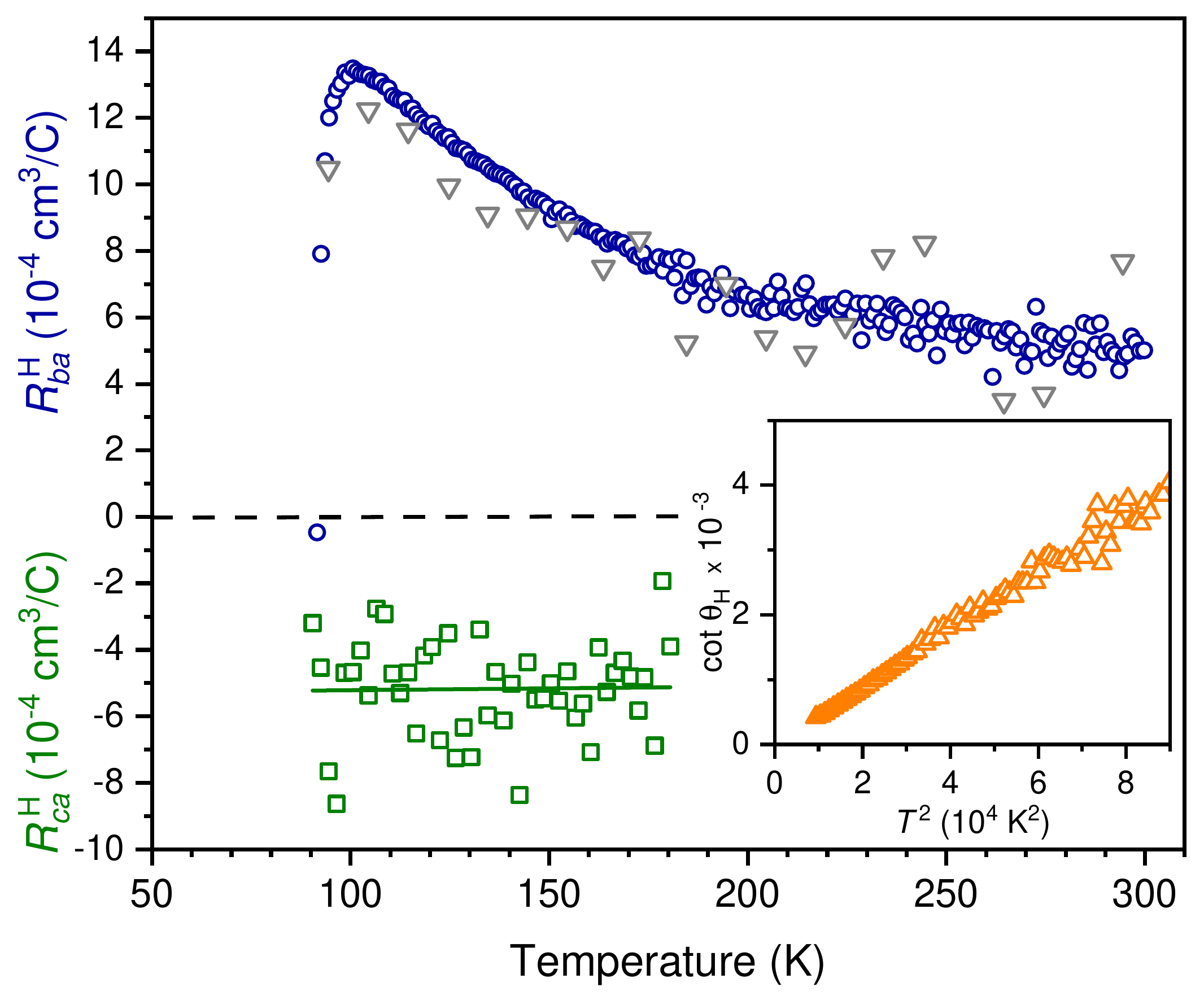}
\caption[]{Temperature dependence of the anisotropic Hall effect in a vicinal YBCO film. The in-plane Hall coefficient $R^H_{ba}$ evaluated from Equation~(\ref{eq:RHbaT}) and averaged to 1\,K intervals is shown by blue circles. Grey triangles represent the alternative evaluation by Equation~(\ref{eq:RHbaL}), averaged to 10\,K intervals. The  out-of-plane Hall coefficient $R^H_{ca}$ (green squares) is determined by Equation~(\ref{eq:RHcaT}) and averaged into 2\,K intervals. The full line is a linear fit to the data. Inset: Cotangent of the in-plane Hall angle $\theta_H$ vs. square of the temperature.} 
\label{fig:Hall}
\end{figure}

The in-plane Hall coefficient $R^H_{ba}$, measured with the magnetic field oriented parallel to the crystallographic $c$ axis and evaluated according to Equation~(\ref{eq:RHbaT}), is presented in \mbox{Figure~\ref{fig:Hall}}. Like in single crystals \cite{TOZE87,HARR92,HOLM93} and $c$-axis oriented thin films \cite{LANG94}, $R^H_{ba}$ of our vicinal YBCO film is positive (hole-like) and increases towards lower temperatures, followed by a sharp drop slightly above $T_c$ as a consequence of the onset of superconducting order parameter fluctuations \cite{LANG94}. A second set of data can be obtained by evaluating \mbox{Equation~(\ref{eq:RHbaL})} yielding consistent results, but with a worse signal to noise ratio (grey triangles).

The inset of Figure~\ref{fig:Hall} confirms that the in-plane Hall angle $\tan \theta_H = R^H_{ba} B_c / \rho_{aa}$ follows Anderson’s law \cite{ANDE91} $\cot \theta_H = A T^2 + C$, where $C$ is proportional to the density of carrier scattering defects and $A$ is a measure of the carrier density. A linear relation between $\cot \theta_H$ and $T^2$ can be observed in the temperature range from $\sim$100 \,K to 300\,K. The intercept C$\sim$0, indicating a minor number of carrier scattering defects---a finding that is also supported by the small intercept of $\rho_{aa}$ discussed above.

A different measurement setup with the magnetic field ${\bf B}=(0,B_b,0)$ oriented parallel to CuO$_2$ planes allows for the evaluation of the out-of-plane Hall coefficient $R^H_{ca}$ according to Equation~(\ref{eq:RHcaT}). In sharp contrast to the positive in-plane Hall effect, the out-of-plane Hall coefficient $R^H_{ca} \sim -5 \times 10^{-4}\, \text{cm}^3/\text{C}$ is negative and almost temperature independent, as can be seen in Figure~\ref{fig:Hall}. The straight line is a linear fit to the data, which does not reveal any temperature dependence, although a slight reduction of $R^H_{ca}$ at lower temperatures cannot be ruled out due to the scatter of data points. Above 180\,K the noise was too large to provide reliable data. The alternative evaluation of $R^H_{ca}$ by Equation (\ref{eq:RHcaL}) confirms its negative sign but was not accurate enough to allow for a quantitative statement. Using films with a larger vicinal angle $\alpha$ would increase the signal-to-noise ratio on the cost of a less regular growth of the YBCO layers, presumably counteracting the advantages of \mbox{larger signals}.

Although the apparent discrepancy of the anisotropic Hall effect in YBCO to conventional transport theory has spurred intense discussions, only few measurements in YBCO single crystals are available. In this respect our investigations on a vicinal film provide a complementary and entirely different measurement method. Our findings for the out-of-plane Hall effect are in good accordance with earlier results on YBCO single crystals, which indicated a slightly temperature dependent $R^H_{ca} \sim -7.5 \times 10^{-4}\ \text{to} -9 \times 10^{-4}\,\text{cm}^3/\text{C}$~\cite{TOZE87}, a temperature independent $R^H_{ca} \sim -6.2 \times 10^{-4}\,\text{cm}^3/\text{C}$ between 100\,K and 400\,K \cite{HARR92}, and $R^H_{ca} \sim -4.2 \times 10^{-4}\,\text{cm}^3/\text{C}$ at 100\,K \cite{ELTS98}.

%%%%%%%%%%%%%%%%%%%%%%%%%%%%%%%%%%%%%%%%%%
\section{Conclusions}

Thin films of YBCO, grown on vicinal-cut SrTiO$_3$ substrates, show anisotropic transport properties when the current is injected either parallel or perpendicular to the growth terraces. From these data, the resistivities parallel and orthogonal to the CuO$_2$ layers can be calculated. The results are in excellent agreement with data in single crystals.

By applying a magnetic field either perpendicular or parallel to the CuO$_2$ layers the in-plane and the  out-of-plane Hall effect can be obtained, respectively. The results are in excellent agreement with single crystal data and confirm the puzzling disparity of in-plane and out-of-plane Hall coefficients regarding their sign and temperature dependence.

We conclude that magneto-transport measurements in vicinal films of layered materials yield equivalent results to measurements in single crystals. The former offer several advantages, like a better controllable sample geometry and a well-defined homogeneous current density. In this respect, vicinal thin films of various superconductors, from copper-oxide to iron-based compounds, might be helpful to study their anisotropic transport properties. Moreover, vicinal thin films are useful for investigations, where a limited penetration depth of photons or ions with moderate energy precludes the use of \mbox{single crystals}.

\vspace{6pt}

%%%%%%%%%%%%%%%%%%%%%%%%%%%%%%%%%%%%%%%%%%
\authorcontributions{W.L. and J.D.P. conceived and supervised the experiments, R.R. and J.D.P. grew and characterized the film, G.H. performed the transport measurements and the analysis, all authors discussed the results, W.L. wrote the paper with substantial contributions from G.H. All authors have read and agreed to the published version of the manuscript.}

\funding{This research was funded by the Austrian Science Fund (FWF) under grant I4865-N and the COST Actions CA16218 (NANOCOHYBRI) and CA19108 (Hi-SCALE) of the European Cooperation in Science and Technology.}

\dataavailability{The data presented in this study are available on reasonable request from the corresponding author.}

\conflictsofinterest{The authors declare no conflict of interest.}

\end{paracol}

\reftitle{References}

%\externalbibliography{no}

\end{document}